\documentstyle[sprocl]{article}
\def\be{\begin{equation}}
\def\ee{\end{equation}}
\def\bea{\begin{eqnarray}}
\def\eea{\end{eqnarray}}

\begin{document}
\title{A  GRAND UNIFIED MODEL WITH\\ $M_G\sim M_{\rm
string}$ AND
$M_I\sim 10^{12}$ GeV}
\author{N. G. DESHPANDE, B. DUTTA}
\address{Institute of Theoretical Science, University
of Oregon, Eugene, OR 97403\\ Preprint \# : OITS-606}
\author{E. KEITH}
\address{Department of Physics, University of
California, Riverside, CA 92521\\ Preprint \# : UCRHEP-T168}
\maketitle\abstracts{
We present a model  based on the gauge group
SU(2)$_L\times$SU(2)$_R\times$SU(4)$_C$ with  gauge couplings that are found to
be   unified at a scale   near the string unification scale. This model
breaks to the MSSM at an intermediate scale which is instrumental in producing
a neutrino in a mass range that can serve as hot dark matter and can also solve
the strong CP problem via a harmless invisible axion.} 

The conventional scale of  supersymmetric grand unification is taken
to be
$M_G\sim 2\cdot10^{16}$ GeV, because this is where the MSSM gauge couplings are
found to converge if one assumes a ``desert" between  about 1 TeV and   and
that scale. However, in superstring theory the unification point  is not  a free
parameter but is predicted to be a function of the gauge coupling at  that scale
in the $\overline{\rm MS}$ scheme as follows \cite{[kap]}:
\begin{eqnarray} M_{\rm string}\approx 7 g_{\rm string}\cdot 10^{17}\, {\rm
GeV}\, ,
\end{eqnarray} which predicts $M_{\rm string}$ to be approximately 25 times
greater  than  the conventional value of $M_G$.  Most possible means of closing
the gap between
$M_G$ and
$M_{\rm string}$ have not yet proven successful in  realistic string models
\cite{[die]}.  However, it has been shown that extra non-MSSM matter that
appears in some realistic  string models can lead to a successful raising of the
 unification scale \cite{[die]}. One obvious and attractive approach to adding
extra matter, ``populating the desert,"  would be to add an intermediate gauge
symmetry. Such realistic string models  have  been built for cases where the
intermediate symmetry is SU(2)$_L\times$SU(2)$_R\times$SU(4)$_C$
\cite{[AL],[AM]}.  Sometimes the field content of these models
have been found to alleviate the discrepancy between the string and gauge
unification scales \cite{[AM]}. 

An intermediate SU(2)$_R\times$SU(4)$_C$ breaking scale of order $10^{12}$ GeV
is very attractive for  two reasons: (1) if the B-L gauge symmetry is broken at
around $10^{11}$-$10^{12}$ GeV, one can easily  get a neutrino mass in the
interesting range of about 3-10 eV, making it a candidate for the hot dark
matter, and (2) if the strong CP problem is solved via the Peccei-Quinn
(PQ) mechanism, this PQ symmetry is required to be broken approximately within
the above window so that  the axion has properties which are consistent with the
lack of observation up to now and the cosmological constraints. As
for the scale  at which the hypothetical PQ-symmetry is broken, perhaps the most
elegant possibility is if it is tied in with the breaking of an intermediate
gauge symmetry, so that there is only one scale between the weak and string
scale to be explained. To obtain the
$\tau$-neutrino mass in the interesting eV range without an intermediate gauge 
symmetry breaking scale
 one has to use a method that either involves a carefully chosen Yukawa coupling
to an
 SU(2)$_R$ triplet, which only arises for particular non-standard Kac-Moody
levels, or non-renormalizable operators with SU(2)$_R$ doublets
\cite{[ML10]}.
 Unlike in the case of an intermediate scale \cite{[edm]}, both these methods
require abandoning the attractive
$b-\tau$ unification hypothesis except in the case of the SU(2)$_R$ doublets
and  high
$\tan\beta \sim m_t/m_b$ which requires greater tuning of the Higgs potential
parameters.   

If  we demand an intermediate scale as mentioned  above, the  necessary
relative  changes \cite{[MR]} in the beta functions of the MSSM are given as
follows: 
\begin{eqnarray}
\Delta b_2-\Delta b_1=2\, ,\, \Delta b_3-\Delta b_2=1\, ,
\end{eqnarray} where the hypercharge has been normalized in the standard GUT
manner and  $b_i=-2 \pi \partial\alpha_i^{-1}/\partial\ln\mu
$. The additional field content we choose at the scale $M_I$ is as
follows: the additional vector representation fields necessary to complete the
intermediate scale gauge symmetry, two copies of the chiral fields
$H=(1,2,4)\equiv ({\bar u}_H^c,{\bar d}_H^c,{\bar E}_H^c,{\bar  N}_H^c)$ and
${\bar H}=(1,2,{\bar 4})\equiv ({ u}_H^c,{ d}_H^c,{ E}_H^c,{  N}_H^c)$, and
chiral singlets $S=(1,1,1)$ which are necessary for the right-handed neutrinos
$N^c_i$ to acquire large Majorana masses. We also add a chiral field
$D=(1,1,6)$ to make all the non-MSSM Higgs modes massive along with two copies 
of chiral fields $\Phi = (2,2,1)$, which contain the MSSM Higgs. There are  of
course the usual three MSSM matter generations that include right-handed
neutrinos 
$F=(2,1,4)\equiv (u,d,\nu ,e)$ and ${\bar F}=(1,2,{\bar 4})\equiv
(u^c,d^c,N^c,e^c)$. The SU(2)$_R\times$SU(4)$_C$ gauge symmetry is broken to the
U(1)$_Y\times$SU(3)$_c$ by
$\left< H\right> =\left< {\bar N}_H^c\right> \, ,\, \left< {\bar H}\right>
=\left< N_H^c\right> \sim M_I$. We note that the existence of the field D and S
are crucial to make all the Higgs modes massive. In string derivations,  
exotic representations tend to occur. General constraints on  
non-minimal   field content are given by:
\begin{eqnarray} n_D\geq 1\, ,\, n_D+n_4=(n_\Phi - 1) + {1\over 2}n_2\, ,
\end{eqnarray}    where $n_D$ is the number of copies of fields transforming as
$(1,1,6)$, $n_\Phi$ is the number of fields transforming as $(2,2,1)$, $n_4$ is
the number of copies of $(1,1,4)+(1,1,{\bar 4})$, and $n_2$ is the number of
copies of $(2,1,1)+(1,2,1)$, which are all to be given mass of order $M_I$.

We note that the appearance of the intermediate breaking scale can occur through
a single parameter in the singlet sector of the model. For example, consider an
R symmetry  invariant superpotential $W=\left(\sum_{i,j=1}^{2} \lambda_{ij}
H_i{\bar H}_j -r\right) S_0 +...$,  where $r$ is of order $M_I$ and $S_0$ has no
VEV. It would then be the most natural case for the VEVs acquired by all four
fields to be of similar order. This mechanism can easily be extended to link the
breaking  of a PQ-symmetry  with the breaking of the intermediate gauge 
symmetry. We also note that in the model we are discussing there are no SU(2)$_R$
triplet fields, therefore one must rely on an extended version of the seesaw
mechanism 
 \cite{[LeeMoha]}.

To allow for the possibility of low
$\tan\beta$, one needs the MSSM Higgs doublets $\phi_u$ and $\phi_d$ to not come
primarily from the same bidoublet $\Phi_i$. We want
one linear combination, from the two bidoublets, of down (or up) type Higgs
superfield SU(2)$_L$ doublets to remain massless and the other combination to
have mass of order
$M_I$. This can be accomplished  through  adding to the model a pair of fields
$H_L$ and ${\bar H}_L$ transforming as
$(2,1,4)$ and $(2,1,{\bar 4})$, respectively, and  also increasing the number of
$H,{\bar H}$ pairs to be $N_H=3$ so that  Eqn.(2) is still satisfied, and using
a  modification of a method that has been used in conventional SO(10) 
GUTs\cite{[BaMoha]}.

We note that an interesting phenomenological result of this model is that
lepton flavor violation  and  the electron's EDM  may
be close to experimental limits. For a more detailed discription of this
model, see  Ref. \cite{[edm]}.

\section*{Acknowledgments} We thank K. S. Babu and E. Ma for 
  discussions. We acknowledge Department of
Energy grants DE-FG06-854ER 40224 and DE-FG02-94ER 40837.

\end{document}